# Study of band bending effect in Dye Sensitized Solar Cell through Constant-Current-Discharging Voltage Decay

Xiaoqi Wang and Chuanbing Cai

Physics Department, Shanghai University, Shanghai 200444, China

A measurement method of constant-current-discharging voltage decay is established to characterize the band bending effect in the heterojunction of conducting glass/$TiO_2$ for typical dye-sensitized solar cells. Furthermore, a dark-state electron transport regarding the $TiO_2$ conduction band bending is proposed based upon the viewpoints of thermionic emission mechanism, which suggests an origin of the band bending effect in a theoretical model. This model quantitatively agrees well with our experimental results and indicates that both the Fermi level decay in $TiO_2$ and the potential difference across the heterojunction will lead to the $TiO_2$ conduction band bending downwards.

Authors' E-mail:

shawl.wang@gmail.com (Xiaoqi Wang)

cbcai@shu.edu.cn (Chuanbing Cai)

# 1. Introduction

Dye-sensitized solar cell (DSSC) and quantum dot sensitized solar cell (QDSSC) have drawn increasing attention during the past decades [1-4]. A lot of efforts have been made on the aspect of photoanode, such as the blocking layer [5], the electrode materials [6] and photoanode decoration [7], and so forth. Recently, it is observed that at the interface of photoanode, i.e. the conducting glass(FTO)/$TiO_2$, the conduction band of $TiO_2$ will bend downwards with the quasi-Fermi level of $TiO_2$ varies even when the cell is under illumination [8-10]. This phenomenon implies the FTO/$TiO_2$ as a kind of rectifying contact, which may have an influence on the current subjected to the electric migration process.

It is now believed, however, that the trap-limited diffusive process plays a crucial role in the photovoltaic current and covers the influence of the electric migration process which is considered negligible [10-14]. To address this issue on the band bending effect, the transport measurement is suggested to be achieved in the dark state, in which the term of electron diffusive process vanishes. Nevertheless, in a composite DSSC there are still some problems in distinguishing the band bending effect from the others, the recombination process at the interface of $TiO_2$/electrolyte and the redox reaction at electrolyte/counter electrode, both of which are dominant in the most conventional measurements, e.g. Cyclic Voltammtry (CV) [15] and Electric Impedance Spectrum (EIS) [16].

To overcome such a difficulty, it is required to establish a method which can effectively clarify the band bending effect. Consider that the open-circuit voltage decay (OCVD), a useful method processed in the dark state, is frequently utilized in characterization of the recombination process at the interface of $TiO_2$/electrolyte [17-18]. The $TiO_2$ thin film appears to be an electron reservoir due to the injection of photo-exited electrons under illumination. As the light is cut off in the case of OCVD, these electrons will release and recombine with electrolyte, see the arrow ($R$) in the Fig. 1(a). Consequently, the potential level $E_{FTO}$ of FTO goes down as the Fermi level $E_F$ of $TiO_2$ decays [19]. It can be proposed to apply a small discharge current on the

FTO as the voltage decay, giving rise to less electron leakage from $TiO_2$, see the arrow (-$I$) in the Fig. 1(a). Moreover, the applied current will cause a potential difference between FTO and $TiO_2$, implying the information on the band bending effect. In this article, we firstly propose an effective method named constant current discharging voltage decay (CCDVD) and then model the dark-state transport concerning the band bending effect, so as to characterize the interface of the heterojunction FTO/$TiO_2$, with respect to the understanding of band bending effect observed in DSSC.

## 2. Experimental details

### 2.1 Preparation of the DSSC sample

A 5×5 mm$^2$ monolayer of $TiO_2$ nanoparticles (size ~20 nm) with thickness of about 4 $\mu$m was prepared by screen printing on the F-doped transparent conducting glass (FTO) substrate. And then it was sealed by a 25 μm-thick plastic spacer together with the platinized CE after sensitizing in the dye of N719 for 20 h. The resultant cell was filled with the electrolyte including 0.1 M LiI, 0.1M $I_2$, 0.5 M 4-tert-butyl pyridine, and 0.6 M 2-Dimethyl-3-propylimidazolium iodide in acetonitrile. The photovoltaic performance for the studied DSSC sample is measured by Keithley 2420. An AM1.5 light was provided by a commercial solar simulator (SAN-EI XES-151S) equipped with a 150W Xenon lamp. The short-circuit current density ($J_{SC}$) and the open-circuit voltage ($V_{OC}$) are approximately 9.5 mA/cm$^2$ and 0.67 V, respectively. The efficient achieved is around 4 %. More details for the sample preparation and structure can be found in our previous publications [20].

### 2.2 Principle for the method CCDVD

OCVD is the powerful method that characterizes the dynamical process of recombination in DSSC. There are a lot of work, both in experimental [17] and theoretical studies [18], in this issue. Under illumination, electrons are injected from the photo-excited dyes to $TiO_2$ thin film, giving rise to the shift upwards of the Fermi level ($E_F$). As the light is cut off, the injected electrons will recombine with electrolyte, resulting in the $E_F$ decay [18]. Moreover, the decay of potential level ($E_{FTO}$) in FTO is coincidence with that of $E_F$ in TiO$_2$, because there is no net charge transfer through the FTO/TiO$_2$ interface, seen in the Fig. 1(b).

As for the measurement of CCDVD, a discharge current is applied on the photoanode as the voltage decays, resulting in a potential difference between FTO and TiO$_2$, as shown in the Fig. 1(c). The applied current depends not only on the characterizations of junction interface of FTO/TiO$_2$ but also on the potential difference of $E_{FTO}$ and $E_F$, so that it appears to be a function of $E_{FTO}$, $E_F$ and $\mathcal{P}$, where $\mathcal{P}$ denotes some characteristic parameters for junction interface. Alternatively, the $E_{FTO}$ can also be written as a function **f** by inversely calculation, that is $E_{FTO}=\mathbf{f}(I, E_F, \mathcal{P})$. As the applied current can be set constant $I_C$ and $E_{FTO}$ is measurable, it suggests that once $E_F$ is derived the parameters $\mathcal{P}$ regarding the characteristic information of FTO/TiO$_2$ interface can be clarified.

A validity condition addressed here to ensure the method effective is that the applied discharge current is too lower than recombination current to affect the variation of $E_F$, so that the decay of $E_F$ in the CCDVD can be identical to that shown in the OCVD. Therefore, in the limit of $I_C \rightarrow 0$, there is

$$E_F(I_C, \mathcal{P}, t) = E_F(0, \mathcal{P}, t) \equiv E_{FTO}(0, \mathcal{P}, t) \tag{1}$$

It is noted that OCVD in principle is the special case of CCDVD at condition $I_C=0$. By eliminating the time parameter $t$, one can derive the time independent relationship of $E_{FTO}(I_C, \mathcal{P})$ v.s. $E_F(I_C, \mathcal{P})$ by the experimental results of $E_{FTO}(I_C, \mathcal{P})$ and $E_{FTO}(0, \mathcal{P})$, allowing us to further characterize the junction interface. In the following measurement of CCDVD, the constant current is supplied by Keithley 2420 digital meter and the voltage decay is recorded by an oscilloscope.

*2.3 Measurement of CCDVD*

Figure 2(a) illustrates the CCDVD measurement for a prepared dye-sensitized solar cell. Keithley 2420 meter is parallel-connected with the circuit and services as a constant current source. As a series of discharge current applied, the decays of $E_{FTO}$ are recorded by the oscilloscope, where has been suggested the redox level is grounded, shown in Fig. 2(b). The curves of $E_{FTO}$ decay can be characterized by three stages. i) In first a few microseconds, the high $E_F$ suggests a large enough electron density in $TiO_2$ to maintain such a small discharge current that a decay in $E_{FTO}$ coincides with the variation of $E_F$ identical to the observation in the measurement of OCVD, i.e. $E_{FTO}(I_C) \approx E_F(I_C) = E_{FTO}(0)$. ii) As $E_F$ moves down, the electrons in $TiO_2$ reduces much, and consequently a difference of $E_{FTO}$ and $E_F$ appears on account of preserving the constant discharge current, that indicates a drop of $E_{FTO}$ in the decay measurement. iii) As $E_F$ approximates to 0 V, however, the recombination process gradually goes weaker than the discharging process, so that $E_F$ decay is gradually dominated by the constant current. This stage has to be excluded in our consideration because it is out of the validity condition of CCDVD, which requires the recombination process dominates the $E_F$ decay. Therefore, only the data of $E_{FTO}$ above 0 V will be taken into account in the results discussion.

Figure 3 summarizes the relationship of $E_{FTO}$ vs. $E_F$ plots in a serious of constant current applied from 1$\mu$A to 5 $\mu$A, where $E_F$ is identical to the $E_{FTO}(I_C=0)$ and the experimental data are represented by various symbols. As the Fermi level $E_F$ decreases the measured voltage $E_{FTO}$ first moves down along the line with a slope of 1 and then gradually drops off, that are the stages i) and ii) identical to the observation in the Fig. 2(b). These behaviors appear to be related to the interface properties, and are supposed to give us more important information with respect to the band bending effect.

## 3. Modeling for heterojunction transport of FTO/TiO$_2$

To further understand the band bending effect from the experimental results of CCDVD, one is required to have basic knowledge of the relationship of $E_{FTO}$ vs. $E_F$ for such heterojunction FTO/TiO$_2$, that is the explicit expression for $E_{FTO}$=f($I$, $E_F$, $\mathcal{P}$).

*3.1 Analysis for the conduction band bending*

In general, bands will bend locally when FTO and TiO$_2$ come in contact, because the two Fermi levels of the materials will equilibrate to the same level through a local exchange of charge carriers [21], the same level that is named quasi-equilibrium state. The photoanode FTO/TiO$_2$ is a n$^+$-n type semiconductor heterojunction, and the conduction band edge of TiO$_2$ bends downwards by several tenths electron volts (*eV*) of $\varepsilon$, leading to an accumulation layer [10, 22], shown in the Fig. 4(a). In the case of OCVD, as the Fermi level $E_F$ decreases due to the recombination process, the junction FTO/TiO$_2$ has experienced a series of quasi-equilibrium state, and in which $E_{FTO}$ is always equal to $E_F$ [23]. For each state of quasi-equilibrium, the band edge ($E_B$) will shift downwards and reach a new equilibrium position, indicating a derivation of $E_B$ from its initial value $E_{B0}$, as shown in the Fig. 4(b). In the method of CCDVD, $E_B$ is considered to further bend owing to the potential difference of $E_{FTO}$ and $E_F$ appears, as illustrated in the Fig. 4(c). Therefore, the variation of $E_B$ is suggested to be a function both of the TiO$_2$ Fermi level $E_F$ and of the difference ($E_{FTO}$-$E_F$), with a first-order form,

$$E_B = E_{B0} + \alpha^{-1}\left(E_F - E_{OC}\right) + \beta^{-1}\left(E_{FTO} - E_F\right) \tag{2}$$

where $E_{B0}$ and $E_{OC}$ denote the initial values of $E_B$ and $E_F$ respectively, that is $E_{B0}=E_C+\varepsilon$ and $E_{OC}=-qV_{OC}$. While $\alpha^{-1}$ denotes $\partial E_B/\partial E_F$ and $\beta^{-1}$ the $\partial E_B/\partial E_{FTO}$, indicating what the extent of band bending is as $E_F$ and $E_{FTO}$ varies, respectively, and they are the characteristic parameters of the junction FTO/TiO$_2$ mentioned above.

*3.2 Transport modeling regarding the band bending*

It is generally believed that under illumination electrons are injected in the TiO$_2$ thin film by the excited dye molecules and then diffuse through the photoanode, suggesting that a diffusive term plays a significant role in the current [10-14]. This term, however, vanishes in the measurement of CCDVD because the light is cut off as the voltage decays. Due to the accumulation layer at junction interface, we address the transport issue based upon the viewpoints of thermionic emission model. A number of electrons that accumulate in the accumulation layer will participate in transport from $TiO_2$ to FTO. Meanwhile, a few electrons in the FTO will also by absorbing thermal energies transfer in the opposite direction, as depicted in the Fig. 4(c). The observed net current can then be described by the transmission function and the electron state density, and is expressed by $I=q\cdot[\Gamma_{TF}\cdot N_C\cdot F(E_B,E_F) - \Gamma_{FT}\cdot N_F\cdot F(E_C,E_{FTO})]$, where $q$ is the elementary charge, $N_F$ and $N_C$ are the states of electron density in region of FTO and $TiO_2$, respectively. $\Gamma_{FT}$ is the transmission rate of electrons transfer in the direction from FTO to $TiO_2$, and $\Gamma_{TF}$ denotes the transmission rate of transfer in the opposite way. $F(E)$ is the Boltzmann distribution function $F(E_1,E_2) = \exp[(E_1-E_2)/k_BT]$ [23, 24], the probability that electron appears in energy level $E_1$ as the chemical potential is $E_2$, where $k_B$ and $T$ are the Boltzmann constant and the absolute temperature, respectively. In equilibrium, i.e. $E_{FTO}=E_F$ and $I=0$, one have $\Gamma_{FT}\cdot N_F = \Gamma_{TF}\cdot N_C\cdot\exp[(E_B-E_C)/k_BT]$ [23], so that the current can be rewritten

$$I = q\Gamma_{TF} N_C e^{\left(-\frac{E_F-E_B}{k_BT}\right)}\left[1-e^{\left(-\frac{E_{FTO}-E_F}{k_BT}\right)}\right] \quad (3)$$

Substituting the band edge $E_B$ by Eq. (2), one derives the explicit expression describing the current flowing through the heterojunction of $FTO/TiO_2$.

$$I = q\Gamma_C n_C \exp\left[\frac{E_{FTO}-\left(1-\frac{\beta}{\alpha}\right)E_F}{\beta k_B T}\right] \\ \times\left[1-\exp\left(-\frac{E_{FTO}-E_F}{k_BT}\right)\right] \quad (4)$$

In the Eq. (4), we have defined $\Gamma_C = \Gamma_{eff}\exp(-E_{OC}/\alpha k_BT)$ and $n_c = N_C\cdot\exp[(E_C-E_F)/k_BT]$, Where, $\Gamma_{eff}=\Gamma_{FT}\cdot\exp(\varepsilon/k_BT)$, is an effective transmission rate independent on the variation of $E_{FTO}$ and $E_F$. As the current is set constant $I_C$, the relationship of $E_{FTO}$ and $E_F$ can then be deduced, seen in Eq. (5). By means of the method CCDVD, $E_F$ and $E_{FTO}$ are measurable at conditions of $I_C=0$ and $I_C\neq 0$, respectively, that allows one to derive the parameters $\alpha$ and $\beta$, and to further understand the band bending effect observed in DSSC.

$$E_{FTO} = \beta k_B T \ln \left\{ \frac{I_C/q}{\Gamma_C N_C} \right.$$

$$\left. + \exp\left[ -\frac{(\beta-1)E_{FTO} - \left(\frac{\beta}{\alpha}-1\right)E_F - \beta E_C}{\beta k_B T} \right] \right\}$$

$$+ \left(\beta + 1 - \frac{\beta}{\alpha}\right) E_F - \beta E_C \qquad (5)$$

## 4. Results and discussion

For the purpose of characterization of junction FTO/TiO$_2$, with respect to the understanding of band bending effect, the Eq. (5) is utilized to fitting our experimental data, as plotted by the solid lines in the Fig. 3. The temperatures and Error rate are listed in the table inserted in Fig. 3. Although there is less change in the temperature, it does affect the fitting results a little, that is possibly because the transport model proposed here is based upon the viewpoint of thermionic emission sensitive to the temperature perturbation. For our DSSC samples, it is derived that $\Gamma_{eff} N_C$=4.798×10$^{18}$ cm$^{-3}$·s$^{-1}$, $\alpha$=1.408, $\beta$=10.204, and $E_C$ approximates to -0.86 V. In general, the electron state density in the conduction band of TiO$_2$ approximates 6×10$^{20}$ cm$^{-3}$ [18], so that the effective transmission rate for our FTO/TiO$_2$ interface is around 8×10$^{-3}$ s$^{-1}$.

More importantly, these results reveal the information of the band bending effect; the conduction band of TiO$_2$ will shift downwards by approximately 70% (1/$\alpha$) times of the $E_F$ decrease, and it will also bend by no more than 10% (1/$\beta$) times of the potential difference between $E_{FTO}$ and $E_F$. It can thus be suggested, even under illumination, that the TiO$_2$ conduction band will bend downwards as the Fermi level of TiO$_2$ decreases, just as what observed in the previous experiments of light interference reflection [9, 10]. Although this effect has an influence on the dark-state current as we modeled, it is still covered in the photovoltaic transport measurements, in which the photocurrent depends mainly on the electron diffusive process due to a large concentration gradient of photo-excited electrons.

## 5. Conclusion

In a summary, the method of voltage decay with a constant discharging current is established to study the effect of the conduction band bending observed in the photoanode FTO/TiO$_2$ of DSSC. The method CCDVD, a general version of the open circuit photovoltage decay (OCVD), allows us to derive the relationship of $E_{FTO}$ vs. $E_F$ in the dark state. To further understand the band bending effect, the explicit expression for the heterojunction transport is deduced based upon the viewpoint of thermionic emission model and agrees well with the experimental results, clarifying the conduction band bending of TiO$_2$ depends not only to a significant extent (70%) on the variation Fermi level in TiO$_2$ but also to a small extent (10%) on the potential difference across the junction. It is believed that the present method can be developed as an effective technology to characterize various photoanode heterostructures, giving rise to crucial information hard to be realized by conventional measurements.

**Acknowledgement:**

This work is partly sponsored by the Innovation Funds for Ph.D. Graduates of Shanghai University (2011), china, the Ministry of Science and Technology of China (973 Projects, No. 2011CBA00105), the National Natural Science Foundation of China (No. 11174193) and the Science and Technology Commission of Shanghai Municipality (No. 10dz1203500 and 11dz00300).

**Fig. 1**

(a) Schematic diagram for heterojunction of FTO/TiO$_2$. The leftwards and rightwards arrows denote the direction of electrons flowing. (b) Energy level diagram of FTO/TiO$_2$ in the OCVD measurement. The downwards arrows denote the time evolution. The dashed lines denote the Fermi levels in respective materials. $E_{ele}$ is the redox level of electrolyte. (c) Energy level diagram in the CCDVD measurement.

**Fig. 2**

(a) Schematic diagram for the method of constant-current-discharging voltage decay. (b) The voltage decays of CCDVD with various constant current from 0 to 10 $\mu A$. The inset re-plots the result of OCVD in the time region of 10 seconds.

**Fig. 3**

The relationship of $E_{FTO}$ vs. $E_F$ concluded from Fig. 2. The hollow symbols denote experimental data and the lines are the fitting curves. The inset table lists the temperature used in fitting and the square average root (SAR) of error rate of fitting.

**Fig. 4**

Energy level diagram for the n$^+$-n type junction of FTO-TiO$_2$. (a) in equilibrium state under illumination; (b) quasi-equilibrium state in the case of OCVD; (c) non-equilibrium state in the case of CCDVD. The red arrows denote direction of electrons transfer.

**Figure 1**

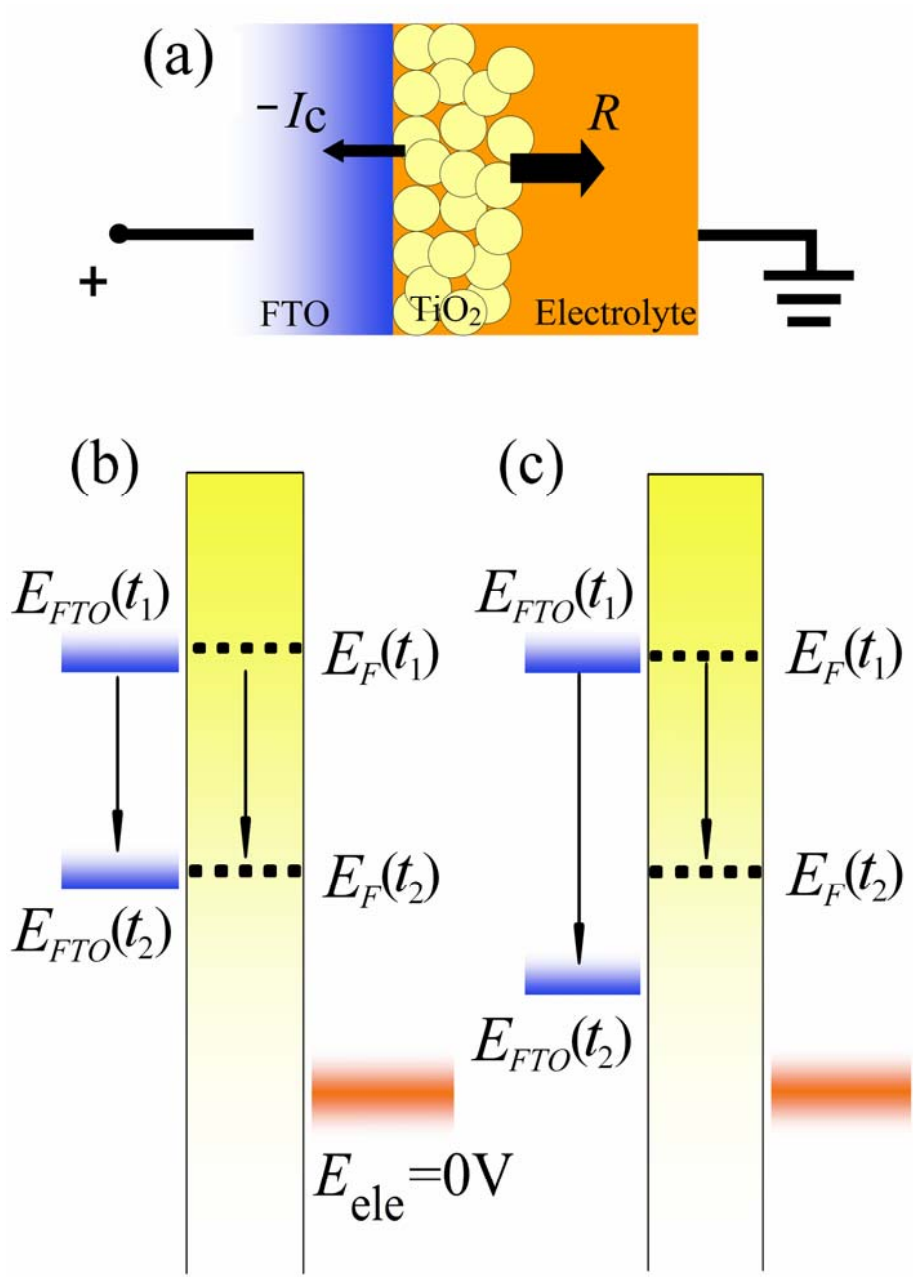

**Figure 2**

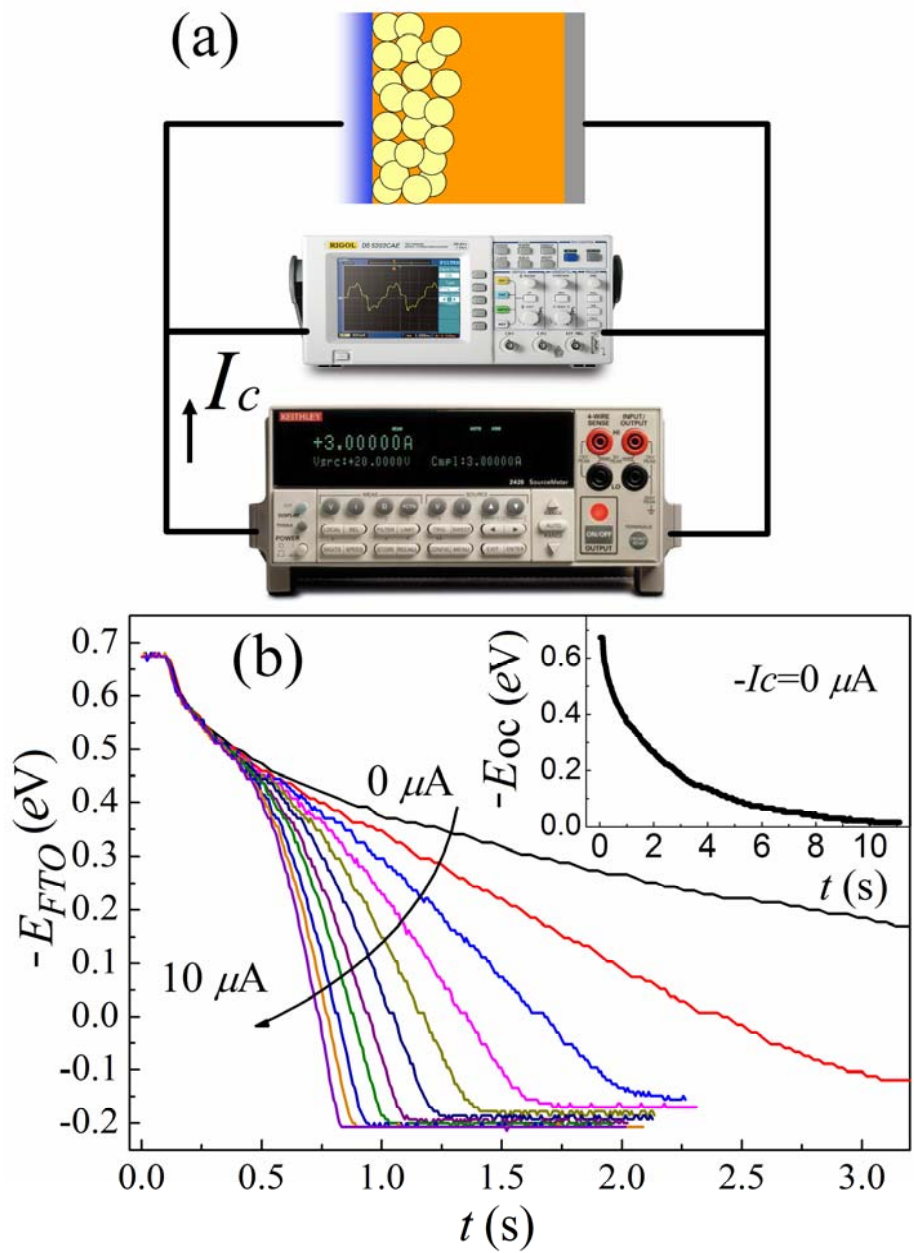

**Figure 3**

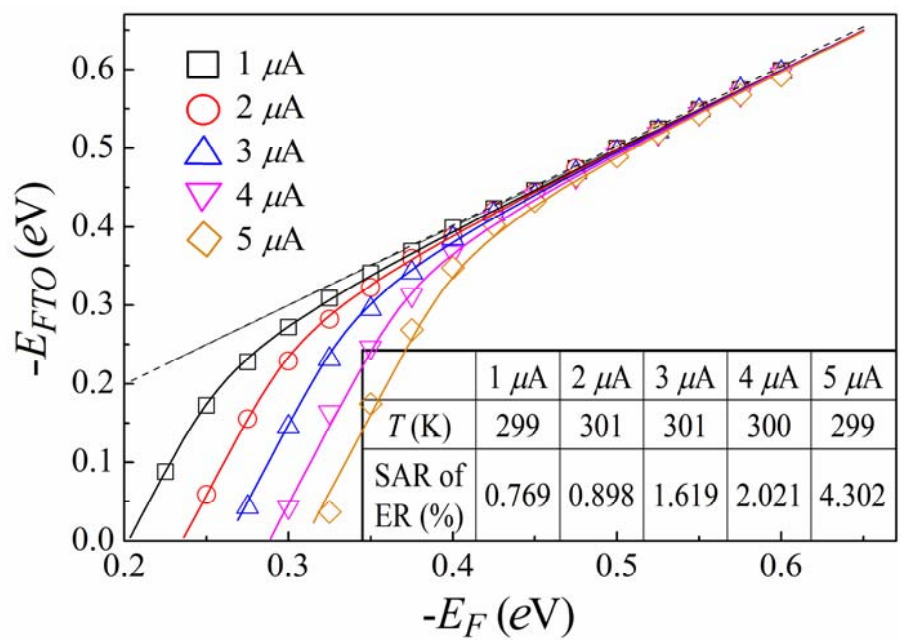

**Figure 4**

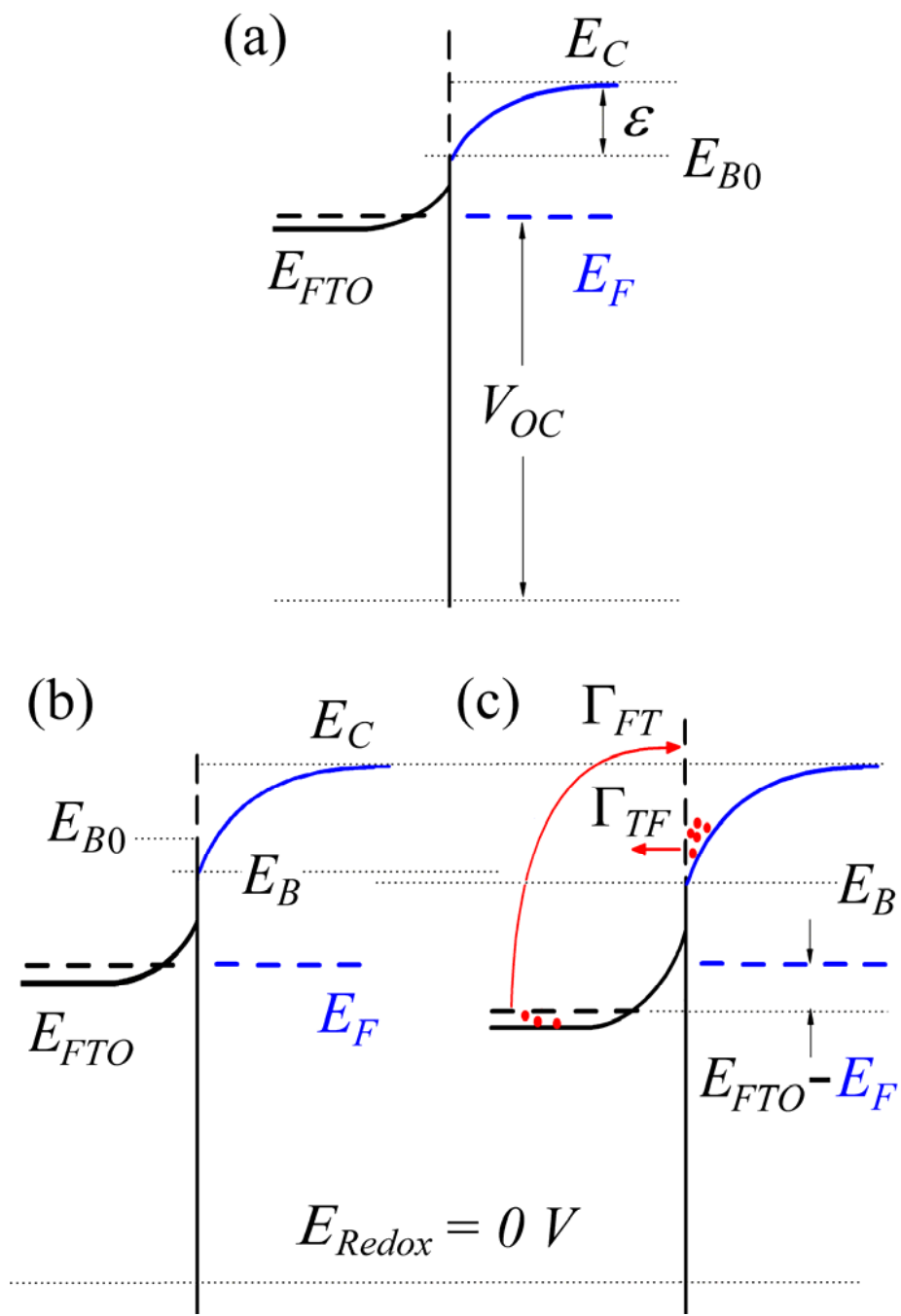